\documentclass[fleqn,usenatbib]{mnras}

\usepackage{newtxtext,newtxmath}

\usepackage[T1]{fontenc}
\usepackage{ae,aecompl}

\usepackage{graphicx}	
\usepackage{amsmath}	
\usepackage[colorinlistoftodos]{todonotes}
\usepackage{hyperref}
\hypersetup{
    colorlinks=true,
    linkcolor=red,
    filecolor=magenta,
    urlcolor=cyan,
}

\usepackage{caption}
\usepackage{parskip}
\usepackage{float}
\usepackage{placeins}
\usepackage{subcaption}
\usepackage{import}
\usetikzlibrary{quotes,arrows.meta}
\usetikzlibrary{positioning}

\usepackage{Ball}
\usepackage{Box}
\usepackage{RightBandedBox}

\usetikzlibrary{positioning}

\newcommand{\xion}{x_{\textsc{Hii}}}

\newcommand{\zmid}{z_{\rm mid}}
\newcommand{\Cion}{C_{\rm ion}}
\newcommand{\Dion}{D_{\rm ion}}

\title[Machine learning from 21 cm maps]{Reproducibility of machine learning analyses of 21 cm reionization maps}

\author[K. Sooknunan et al.]{
K. Sooknunan$^{1}$\thanks{E-mail: k.sooknunan19@imperial.ac.uk (KS)},
E. Chapman$^{2}$,
L. Conaboy$^{2}$,
D.~J.~Mortlock$^{1,3,4}$ and
J.~R.~Pritchard$^{1}$\thanks{E-mail: j.pritchard@imperial.ac.uk (JP)}
\\~\\
$^{1}$Astrophysics Group, Imperial College London, Blackett Laboratory, Prince Consort Road, London SW7 2AZ, UK\\
$^{2}$School of Physics and Astronomy, The University of Nottingham, University Park, Nottingham, NG7 2RD\\
$^{3}$Department of Mathematics, Imperial College London, London, SW7 2AZ, UK\\
$^{4}$Department of Astronomy, Stockholm University, Albanova, SE-10691 Stockholm, Sweden
}

\pubyear{2024}

\begin{document}
\label{firstpage}
\pagerange{\pageref{firstpage}--\pageref{lastpage}}
\maketitle

\begin{abstract}
\noindent
Machine learning (ML) methods have become popular for parameter inference in cosmology, although their reliance on specific training data can cause difficulties when applied across different data sets. By reproducing and testing networks previously used in the field, and applied to 21cmFast and Simfast21 simulations, we show that convolutional neural networks (CNNs) often learn to identify features of individual simulation boxes rather than the underlying physics, limiting their applicability to real observations. We examine the prediction of the neutral fraction and astrophysical parameters from 21~cm maps and find that networks typically fail to generalise to unseen simulations. We explore a number of case studies to highlight factors that improve or degrade network performance. These results emphasise the responsibility on users to ensure ML models are applied correctly in 21~cm cosmology.
\\
\end{abstract}



\graphicspath{{plots/}}


\section{Introduction}
\label{sec:intro}

Our understanding of the Universe has grown substantially over the last few decades, but there is still much we do not know about the Epoch of Reionization (EoR) -- the period of $\sim\!1$ billion years after the Big Bang when light from the first stars ionized the hydrogen in the intergalactic medium (IGM). There are many possible astrophysical models for reionization, typically characterized by parameters such as the escape fraction of ionizing photons from galaxies \citep{Khaire:2016,Mesinger:2016}, and there is considerable uncertainty in the overall timing of the reionization process \citep{Greig:2017b,Kulkarni:2019}. Current empirical constraints about the EoR are limited, coming mainly from cosmic microwave background (CMB) measurements \citep{PlanckCosmo:2020} and quasar absorption spectra \citep{Fan:2006,Becker:2018,Keating:2020,Bosman:2022}.

Recent observations from the James Webb Space Telescope (JWST)\footnote{https://www.stsci.edu/jwst} have provided critical insights into the Epoch of Reionization, particularly in refining the timeline. JWST’s deep-field observations have revealed galaxies forming as early as 200 million years after the Big Bang \citep{Naidu:2022,Finkelstein:2023,Harikane:2023,Curtis-Lake:2023, Carniani:2024}, pushing back the likely start of reionization. Furthermore, estimates from JWST data suggest that the star formation rates and escape fraction of photons might be higher than previously anticipated, especially in faint, early galaxies \citep{Atek:2024}. JWST’s ability to study faint star-forming galaxies \citep{Endsley:2023,Endsley:2024}, constrain and detect the ionized bubbles surrounding galaxies \citep{Saxena:2024,Hsiao:2023}, as well as the measuring of neutral hydrogen fractions via Lyman-alpha absorption damping wings \citep{Umeda:2024}, has been instrumental in mapping the progress of reionization.

The most promising new probe of the EoR is the redshifted 21 cm line of neutral hydrogen \citep[for a review see e.g.][]{Pritchard:2012,Liu:2019}. This is the radiation released by a neutral hydrogen atom when its electron undergoes a spin flip transition \citep{Field:1958,Field:1959b}. A new generation of low-frequency radio telescopes, such as the Low Frequency Array \cite[LOFAR\footnote{www.lofar.org},][]{vanhaarlem:2013}, the Hydrogen Epoch of Reionization Array (HERA\footnote{http://reionization.org}, \cite{HERA}), and the Square Kilometre Array \cite[SKA\footnote{www.skao.int},][]{Braun:2015}, should be able to directly observe the evolution of the neutral hydrogen distribution during the EoR.

An open problem in EoR research is how best to constrain the reionization history using 21 cm intensity measurements. One common approach is to use summary statistics such as the power spectrum \citep{Trott:2016,Pober:2014}; and fully Bayesian parameter inference methods have also been developed for this task \citep{Greig:2015,Greig:2017}. However, the 21~cm signal from the EoR is expected to be highly non-Gaussian, implying that there is significant information encoded in higher-order statistics such as the bispectrum \citep{Majumdar:2018,Shimabukuro:2016}. But all such summaries inevitably entail some loss of information. The SKA will produce tomography throughout the EoR, allowing the possibility of inferring parameters from the maps themselves. However, the amount of data and large parameter space pose significant challenges for traditional statistical inference approaches, motivating the use of more flexible machine learning (ML) methods.

ML methods have already been shown to be effective in the analysis of 21cm simulations. Artificial neural networks (ANNs) have been used for emulation to accelerate the speed of modelling \citep{Schmit:2018,Bevins:2021,Bye:2022,Sikder:2024}, to improve foreground mitigation \citep{Mertens:2024} and for inference of parameters from the power spectrum \citep{Shimabukuro:2017,Choudhury2020,Choudhury2021, Choudhury:2022,Tiwari:2022,Shimabukuro2022}. Convolutional neural networks (CNNs) have been used to emulate summary statistics \citep{Breitman:2024}, infer parameters from power spectra \citep{Doussot:2019}, images \citep{Gillet:2019,Kwon:2020,Neutsch2022,Choudhury:2022} and lightcones \citep{Zhao:2022,Prelogovic:2022}, to infer topological statistics \citep{Bianco:2021}, constrain cosmological models \citep{Sabiu:2022} and for foreground removal \citep{Villanueva:2021,Makinen:2021,Shulei:2022,Bianco:2024,Sabti:2024}.

In this paper, we investigate the reproducibility and generalisability of currently proposed CNN-based methods of reionization parameter inference from 21~cm images.
After briefly summarising 21~cm cosmology in \autoref{sec:back} and CNNs in \autoref{sec:formal}, we then look at several case studies:







\begin{itemize}

\item
In \autoref{sec:mangena_et_al} we create a simulated 21cm signal and reproduce a common CNN methodology used in the field. We first attempt to recover the neutral fraction of hydrogen, before and after generalising the network.

\item 
In \autoref{sec:hassan_et_al} we apply this same network to the simulated data set, but with the aim of recovering a wider range of astrophysical and cosmological parameters: $\Omega_{\mathrm{m}}, h, f_{\mathrm{esc}}, \sigma_8, C_{\mathrm{ion}} \text{ and } D_{\mathrm{ion}}.$

\item In \autoref{sec:5.3} we test the correlation between performance of the network and the number of training examples.

We then look at adding redshift information structure into the network, such that we can recover parameters that capture the evolutionary or integrated nature of reionization. 
\item
In \autoref{sec:laplante_et_al} we use the network to constrain the timescale and duration of reionization.

\item
In \autoref{sec:billings_et_al} we attempt to infer the CMB optical depth from sets of 21cm maps at different redshifts. 

\end{itemize}

In \autoref{sec:sixparams} we explore whether the network can be improved by including additional redshift information, and summarise our conclusions in \autoref{sec:conc}.



\section{21~cm cosmology}
\label{sec:back}

Hydrogen is the most abundant element in the Universe. The 21~cm spin-flip transition of neutral hydrogen atoms \citep{Field:1958,Field:1959b}, with an observed frequency, for a source at redshift $z$, of $\nu_{\rm obs}=1420 {\rm\, MHz}/(1+z)$, is detectable by radio telecopes for the entirety of the EoR. The distribution of neutral hydrogen is expected to be sensitive to various cosmological parameters, facilitating the emerging field of 21~cm cosmology. The key observable is differential brightness temperature (\autoref{sec:dbt}), which provides information both along the line-of-sight and across the sky. It is, however, challenging to predict the expected signal, so 21~cm cosmology relies on fast reionization simulation methods (\autoref{sec:simfast21}). The impossibility of specifying a conventional likelihood for 21~cm data means that model predictions can only be represented via large numbers of simulations  (\autoref{sec:simfast21_dataset}) which we then use to train and test ML parameter estimation methods (\autoref{sec:formal}).


\subsection{Differential brightness temperature}
\label{sec:dbt}

The differential brightness temperature of the redshifted 21 cm line seen at frequency $\nu$ along a particular line-of-sight is given by \cite[e.g.][]{Madau:1997}
\begin{align}
\delta T_{b}(\nu) 
                  \simeq & \, 27 \, x_{\mathrm{HI}} \, \left(1+\delta\right) \, \frac{1}{1 + ({\rm d} v_{r} / {\rm d} r) / H(z)} \nonumber\\ 
                  &\times \, 
                  \left(1-\frac{T_{\rm CMB}}{T_{\mathrm{S}}}\right) \,
                  \left(\frac{1+z}{10} \frac{0.15}{\Omega_{\mathrm{m}} \, h^{2}}\right)^{1 / 2}\left(\frac{\Omega_{\rm b} \, h^{2}}{0.023}\right) \, \mathrm{mK}\;\;,
\label{eq:delta_t}
\end{align} 
where $T_S$ is the spin temperature, $T_{\rm CMB}$ is the CMB temperature, $x_{\rm HI}$ is the neutral fraction of hydrogen, $\delta = \delta(\mathbf{x},z) \equiv \rho/\bar{\rho} - 1$ is the evolved Eulerian density contrast, $v_{r}$ is the comoving velocity and $H(z)$ is the Hubble parameter at redshift $z$. This is given as 
\begin{equation}
\label{eq:H}
H^{2}(z)=H_{0}^{2}\left[\Omega_{\rm m}(1+z)^{3}+\Omega_{\rm r}(1+z)^{4}+\Omega_{\Lambda}\right] ,
\end{equation}
where $\Omega_{\rm m}$, $\Omega_{\rm r}$ and $\Omega_{\Lambda}$ are the normalised matter, radiation and dark energy densities respectively. \autoref{eq:delta_t} is valid under the assumption that ${\rm d} v_{r} / {\rm d} r \ll H$. 

In this paper, we will work in the saturated regime $T_S\gg T_{\rm CMB}$, where the dependence on spin temperature can be ignored. Strictly speaking this is unlikely to be true for redshifts $z\gtrsim8$ \citep{Furlanetto:2006}, but the details are not important for our main purpose of testing the robustness of neural network architectures. We defer to future work the inclusion of spin-temperature fluctuations.

\subsection{Simulation of the 21~cm signal}
\label{sec:simfast21}

To calculate the 21~cm signal from the EoR it is necessary to simultaneously model the large-scale density field and the small-scale propagation of photons that ionise the IGM, requiring simulations with large dynamic range. 

One approach is to use a full radiative transfer (RT) simulation in which the calculation is split into two parts. First, the large scale structure of the Universe is simulated using $N$-body simulations. Then the radiation field due to ionising sources is treated using detailed radiative transfer and non-equilibrium simulations \citep[e.g.][]{Iliev:2015, Iliev:2006, Mellema:2006, McQuinn:2007, Shin:2008, Baek:2009, Thomas:2009, Ghara:2015}. This takes all the underlying physics into account, making RT simulations very accurate; however, this accuracy comes at a significant computational cost. 

One can reduce this computational cost using semi-numerical methods \citep[e.g.][]{Zahn:2007, Mesinger:2007,
Santos:2008, Choudhury:2009}. These methods use analytic approximations instead of full $N$-body simulations, but calculate the brightness temperature fields numerically using the ionization field (discussed further below). While these methods are approximate, they have a much lower computational cost and have been shown to give acceptable agreement with RT simulations \citep[e.g.][]{Zahn_2011,Hutter2018,Molaro:2019}.

In this work we adopt the semi-numerical approach as implemented in \texttt{SIMFAST21}\footnote{https://github.com/mariogrs/Simfast21} \citep{Santos:2008,Hassan:2016}. \texttt{SIMFAST21} simulates the 21~cm background at high redshifts and computes the difference in brightness temperature between the background and the CMB as given in \autoref{eq:delta_t}.

\begin{figure*}
     \centering
    \includegraphics[width = \linewidth]{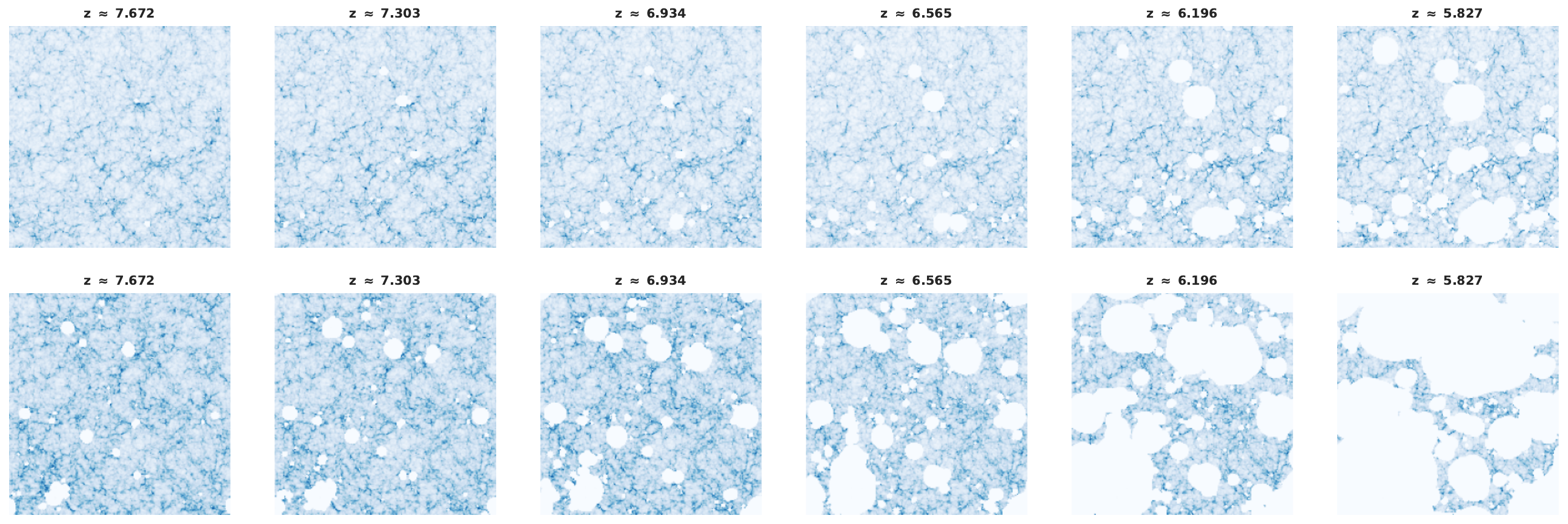}
    \caption[]{Brightness temperature fields for two example \texttt{SIMFAST21} simulation boxes of size of 150~Mpc, going from redshift $z = 7.7$ (left) to $z = 5.8$ (right). The top row shows the evolution for a late model; the bottom row shows the evolution for an early model.
    }
    \label{fig:simfast21sim}
\end{figure*}

\texttt{SIMFAST21} starts by generating a linear matter density field using a random Gaussian distribution. These density fields are used to identify the locations of collapsed halos using the excursion set formalism \citep[see][for details]{Santos:2008}. A region is considered to undergo gravitational collapse if $\bar{\delta}(M,z) > \delta_c(M,z) $, where $\bar{\delta}(M,z)$ is the mean overdensity of the region and $\delta_c(M,z)$ is the critical overdensity required for collapse. A given cell is considered to be the centre of a halo of mass $M$ if it meets the criteria outlined above. By repeating this for all cells, a catalogue of halo masses and locations is created. The locations of the dark matter halos are then corrected using a linear velocity field evolved following the \cite{Zeldovich:1970} approximation. 

The galaxies that form in these collapsed dark matter halos emit photons that ionise the IGM, for which the rate, $\Gamma$, is assumed to be related to the halo mass by
\begin{equation}
\label{eq:Rion}
\frac{R_{\rm ion}}{M} = A(1+z)^{D_{\rm ion}} \, \left(\frac{M}{B} \right)^{C_{\rm ion}} \,  {\rm e}^{- (B / M)^3}.
\end{equation}
Fiducially, 
$A = 1.08\times10^{40} \, {\rm M}_{\odot}^{-1}\, {\rm s}^{-1}$, 
$B = 9.51\times10^{7} \, {\rm M}_{\odot}$, $C_{\rm ion} = 0.41$, $D_{\rm ion} = 2.28$.
Ionised regions can be identified by applying the excursion set formalism to these corrected halo mass catalogues with the assumption that regions self-ionize \citep{FZH:2004}. A region of radius, $R$, is considered to be ionised if $f_{\rm coll} \geqslant \zeta^{-1}$, where  $f_{\rm coll}$ is the fraction of mass that has collapsed in halos in that region and $\zeta$ is some efficiency parameter. Starting off with a large $R$ and moving to $R = R_{\rm cell}$, if the region meets the criteria then all cells in the radius $R$ will be marked as ionised and assigned a value of 1. If the criteria was not met at $R = R_{\rm cell}$, then the cell will be assigned the value which is equal to the ratio between the ionised volume and cell volume. This approach efficiently calculates an ionizing field. Once we have the ionization field and the density field, we can then calculate the brightness temperature using \autoref{eq:delta_t}. 

\autoref{fig:simfast21sim} shows the brightness temperature fields for an example \texttt{SIMFAST21} simulation and illustrates the evolving structure that can, if recovered to a high accuracy, enable 21 cm cosmology.

\subsection{Simulation dataset}
\label{sec:simfast21_dataset}

We created $30,000$ reionization simulations from \texttt{SIMFAST21} using a different random initial seed for each simulation so that each had a unique density field. Each simulation had 21 boxes, sampled at equally spaced redshifts from $z = 5$ to $z = 12$. Each box had $200^3$ cells and a (co-moving) side length of $150$ Mpc.

For each simulation, we use Latin hypercube sampling to explore these six parameters (with parameter ranges in brackets):
\begin{itemize}
    \item 
    normalised matter density, $\Omega_{\rm m}$: $[0.2,0.4]$
    \item 
    dimensionless Hubble constant, $h$: $[0.6,0.8]$
    \item 
    matter fluctuations amplitude, $\sigma_8$: $[0.7,0.9]$
    \item 
    photon escape fraction, $f_{\rm esc}$: $[0.01,1]$
    \item 
    the $\Gamma_{\rm ion}$-$M_{\rm h}$ power-law dependence, $C_{\rm ion}$: $[0,1]$
    \item 
    $\Gamma_{\rm ion}$ redshift evolution index, $D_{\rm ion}$: $[0,2]$
\end{itemize}

The first four are standard in 21~cm cosmology parameters but $C_{\rm ion}$ and $D_{\rm ion}$ are particular to \texttt{SIMFAST21} and its ionization rate parameterisation (\autoref{eq:Rion}).  We made minor adjustments to the code in order to explore different values of $C_{\rm ion}$ and $D_{\rm ion}$.

With this large suite of reionization simulations in hand we are now in a position to be able to train ML models to extract the underlying parameters from the boxes.  



\section{Parameter estimation using ML methods}
\label{sec:formal}

As detailed in \autoref{sec:intro}, there is no practical way of evaluating the likelihood for a full dataset, so likelihood-based approaches with restricted summary statistics inevitably lose information. Thus, conventional statistical techniques are of limited use for 21~cm cosmology. This motivates the use of ML methods which utilise  simulated datasets of the type described in \autoref{sec:back}. The versatility of ML means that models can be trained to learn the relationship between observables and underlying model parameters reliably enough that they can be used to infer parameter values from unseen data. In 21~cm cosmology, this type of problem has already been tackled with ANNs (\autoref{sec:ann}) with partial success. This is the approach we adopt; our model architecture is described in \autoref{sec:network}.





\subsection{Artificial neural networks}
\label{sec:ann}


The default ML method for image analysis is ANNs \citep{Simonyan:2014}, in which the input data is connected through layers of neurons to a (typically smaller) number of outputs. The functions linking neurons (also refered to as nodes) in adjoining layers are typically simple in form, with a common option being the rectified linear unit (ReLU) activation, in which the output for an input node of value $x$ is  $f(x) = \textrm{max}(0,x)$. The links between layers can either connect all pairs of nodes in a fully connected layer, or only nearby nodes in, e.g., a convolutional layer \citep{LeCun:1990,LeCun:1998}. The overall result is an arbitrarily complicated function which, constrained by sufficient training data, can then be used to encode the appropriate mapping between data and parameters. The number of layers, the number of neurons in each layer and the type of the layers/connections is defined as the architecture of the network, and is the main design choice in any given ML problem. Common options include: deep networks \citep{Schmidhuber:2014,Goodfellow:2016}, in which there are many hidden layers between the input and output; CNNs in which at least one of the connections is a convolution \citep{LeCun:1990}; and the use of 
max pooling layers in which the input is downsampled to a smaller number of outputs. 

The choice of architecture depends on the nature of the data and the problem; for example, CNNs are often preferred for image data due to their ability to efficiently capture spatial hierarchies. Selecting the right network is crucial, as a poorly matched architecture can lead to suboptimal performance or overfitting. A well-designed network ensures that the model effectively captures the underlying structure of the data while being generalisable to new datasets.

\subsection{Our network architecture}
\label{sec:network}

\begin{figure*}
    \centering
    \includegraphics[width = \linewidth]{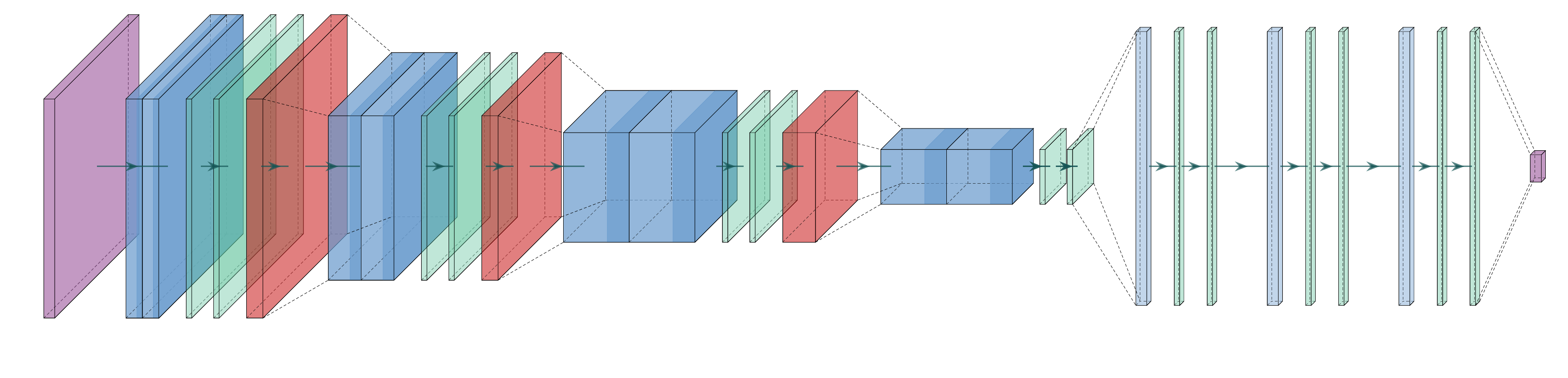}
  \caption{Our network architecture, for more details on each layer see \autoref{tab:arch}. The network consists of two sections: a convolutional section and a fully connected section. The convolutional section has four convolutional blocks, each with two convolutional layers, followed by batch normalisation and a ReLU activation, with max pooling layers in between. The fully connected section has three blocks, each with a fully connected layer, batch normalisation, and a ReLU activation. 
  }
  \label{fig:arch}
\end{figure*}

The network architecture chosen for this work is one of the CNN networks (Network I) presented in \cite{Hassan:2019}. This network is specified in \autoref{tab:arch} and illustrated in \autoref{fig:arch}.  This network was shown to produce good parameter recovery with relatively few training epochs. The network can be split into two sections: a convolution section; and a fully connected section. The first section comprises of four convolutional blocks, joined together with max pooling layers. Each convolutional block is made up of two convolutional layers, followed by a batch normalisation layer, and a ReLU activation layer. The second section comprises of three fully connected blocks.  Each fully connected block is made up of a fully connected layer, followed by a batch normalisation layer and a ReLU activation layer.  Biases were turned off for all layers, following \cite{Hassan:2019}, and the padding for the convolution layers was chosen so that the input and output have the same shape, as is standard for recovering images. 

The parameters used to train our network are listed in \autoref{tab:train_param}. A Glorot (or Xavier) initialisation \citep{glorot:2010} was used for the weights of all layers, which were drawn from a normal distribution of mean $\mu=0$ and variance $\sigma^2 = 2 / (N_{\rm input} + N_{\rm output})$, i.e., the variance is the inverse mean of the number of input neurons, $N_{\rm input}$ and output neurons, $N_{\rm output}$. This ensures that the variance in the input maps is preserved as it moves through the layers of the network.

We used a mean squared error (MSE) loss function, defined as
\begin{equation}
\mathrm{MSE}=\frac{1}{n} \sum_{i=1}^{n}(y_{{\rm true},i}-y_{{\rm pred},i})^{2},
\end{equation}
where $n$ is the number of data points and $y_{{\rm true},i}$ and $y_{{\rm pred},i}$ are the true and predicted values in the $i^{\rm th}$ cell.

We used the adaptive moment estimation (Adam) algorithm \citep{kingma:2014} to update the network weights during the training process, with updates being made in small increments controlled by a learning rate of $0.001$. The networks were trained with a batch size of 128 for between 100 and 200 epochs, where an epoch is defined as one full pass through of the training set. The network training step took $\sim\!15$ hours using the Imperial College HPC with a Tesla K40c GPU.

To quantify the performance we use the $R^2$ score, defined as \citep{Hassan:2019}
\begin{equation}
    R^2 = 1 - \frac{\sum_{i=1}^n (y_{{\rm true},i} - y_{{\rm pred},i})^2}{\sum_{i=1}^n (y_{{\rm true},i} - \bar{y}_{{\rm true},i})^2}\;,
\end{equation}
where $\bar{y}_{\rm true}$ is the average of all the labels actually in the test sample.

\begin{table}
\caption{Network architecture specifications}
\begin{tabular}{llp{1.1cm}c}
\hline
\textbf{layer} & \textbf{parameters}                               & \textbf{output shape} \\
\hline
input          & (none)                                              & (200,200,1)           \\
conv2D         & filters=32, kernel=(3,3), strides=(1,1)$\!\!\!\!$  & (200,200,32)          \\
conv2D         & filters=32, kernel=(3,3), strides=(1,1)$\!\!\!\!$  & (200,200,32)          \\
batch norm     & (none)                                              & (200,200,32)          \\
activation     & ReLU                                              & (200,200,32)          \\
max pool.\ 2D$\!\!\!\!$ & pool size=(2,2), strides=(2,2)                & (100,100,32)          \\
conv2D         & filters=64, kernel=(3,3), strides=(1,1)$\!\!\!\!$  & (100,100,64)          \\
conv2D         & filters=64, kernel=(3,3), strides=(1,1)$\!\!\!\!$  & (100,100,64)          \\
batch norm     & (none)                                              & (100,100,64)          \\
activation     & ReLU                                              & (100,100,64)          \\
max pool.\ 2D$\!\!\!\!$ & pool size=(2,2), strides=(2,2)                & (50,50,64)            \\
conv2D         & filters=128, kernel=(3,3), strides=(1,1)$\!\!\!\!$ & (50,50,128)           \\
conv2D         & filters=128, kernel=(3,3), strides=(1,1)$\!\!\!\!$ & (50,50,128)           \\
batch norm     & (none)                                              & (50,50,128)           \\
activation     & ReLU                                              & (50,50,128)           \\
max pool.\ 2D$\!\!\!\!$ & pool size = (2,2), strides=(2,2)                & (25,25,128)           \\
conv2D         & filters=256, kernel=(3,3), strides=(1,1)$\!\!\!\!$ & (25,25,256)           \\
conv2D         & filters=256, kernel=(3,3), strides=(1,1)$\!\!\!\!$ & (25,25,256)           \\
batch norm     & (none)                                              & (25,25,256)           \\
activation     & ReLU                                              & (25,25,256)           \\
dense          & neurons = 1024                                    & (1024)                \\
batch norm     & (none)                                              & (1024)                \\
activation     & ReLU                                              & (1024)                \\
Dense          & neurons = 1024                                    & (1024)                \\
batch norm     & (none)                                              & (1024)                \\
activation     & ReLU                                              & (1024)                \\
dense          & neurons = 1024                                    & (1024)                \\
batch norm     & (none)                                              & (1024)                \\
activation     & ReLU                                              & (1024)                \\
dense          & neurons = 6                                       & (6)             \\
\hline     
\end{tabular}
\label{tab:arch}
\end{table}

\begin{center}
\begin{table}
\caption{Parameters used to train the network}
\begin{tabular}{ll}
\hline
\textbf{parameter} & \textbf{value}     \\
\hline
loss               & mean squared error \\
optimiser          & Adam               \\
learning rate      & 0.001              \\
metric             & root mean squared  \\
batch size         & 128             \\
\hline
\end{tabular}
\label{tab:train_param}
\end{table}
\end{center}

We now turn to applying these techniques to four different case studies drawn from the literature \citep{Hassan:2019,Mangena:2020,LaPlante:2018,Billings:2021} to investigate the generalisability of CNNs when applied to 21 cm maps. These case studies have been selected as representative of ways that CNNs have been applied to infer different parameters from simulated 21cm maps. 

\section{Recovering the Ionization Fraction}
\label{sec:mangena_et_al}
We first attempt to use a CNN trained on redshift slices from 21cm simulations to constrain the ionization fraction, $\xion$, of an individual input 21 cm image. \cite{Mangena:2020} found that CNNs were able to determine the ionization fraction with a 99\% accuracy. They made use of a training set that drew slices from 500 Mpc boxes with 200$^3$ pixels, covering redshifts $z=$ 6-10. We follow their approach here, using our own simulations to explore its ability to generalise to unseen data. Our approach mirrors theirs with two main differences: we use smaller boxes (150 Mpc rather than 500 Mpc), and we take steps to produce a balanced training set as described below. These changes enable a more consistent comparison across all the later case studies. 

\citet{Iliev:2014} showed that 150 Mpc simulation volumes are large enough to ensure large-scale 21cm fluctuations are captured to the extent that reionization histories converge, and we achieve the same degree of parameter recovery with the smaller volume. 

\subsection{Data pre-processing}
\label{sec:preproc_1}
\texttt{SimFast21} outputs simulation cubes at each redshift from which we extract individual slices to form the training set. In addition, we must ensure that the data set given to the CNN is balanced. For example, if given a data set dominated by slices of high ionization fraction, the network would learn this bias and favour specific reionization histories with that bias. We mitigate this by selecting a subset of slices which gives a more uniform distribution of ionization fraction across the slices. 

\begin{figure}
\centering
\includegraphics[width = \columnwidth]{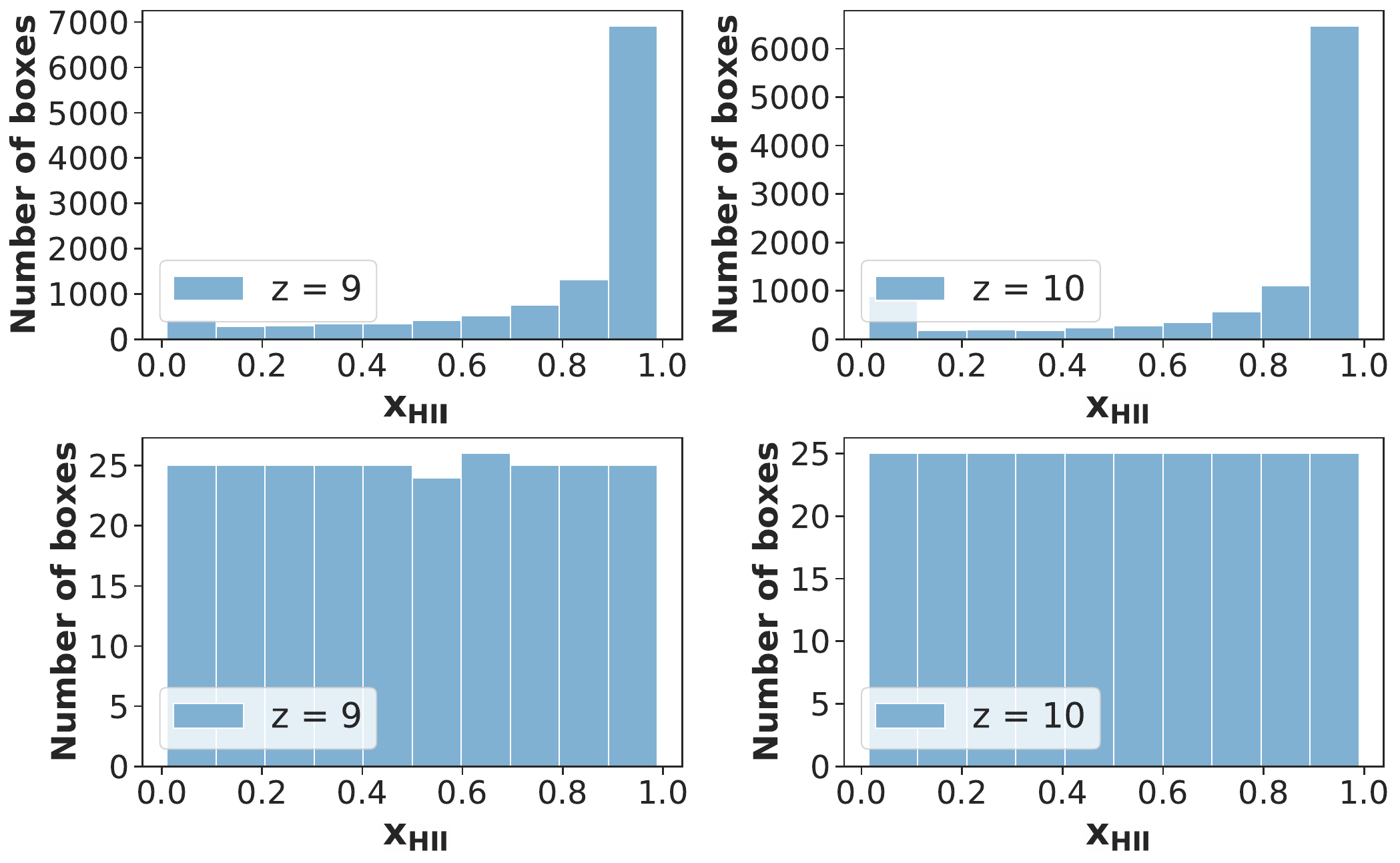}
\caption{Histograms of the ionization fraction ($\xion$) distribution of our dataset at redshifts of $z = 9$ and $z = 10$. The dataset is very skewed towards higher ionization fractions (top panel); this is fixed by balancing the dataset (bottom panel) such that equal numbers of slices are drawn from each ionization fraction bin. }
\label{fig:unbal_dist}
\end{figure}

The required pre-processing steps are:
\begin{enumerate}

     \item 
     Remove all simulation cubes that have an ionization fraction, $\xion<0.01$ or $\xion>0.99$. Brightness temperature maps with these $\xion$ values are featureless---they have essentially constant values in all pixels---despite having drastically different $\xion$ values. If these were left in, the network would be unable to distinguish between the two scenarios leading to confusion.
     
     \item 
     The top panel of \autoref{fig:unbal_dist} shows the distribution of $\xion$ for our simulation boxes and demonstrates that the number of boxes increases as $\xion$ increases. This is because our parameter space favours simulation boxes with larger $\xion$: many combinations of allowed parameters lead to near complete ionization of the boxes. This unbalanced data set would cause our training set to be biased. To normalise this distribution, we randomly sample 25 simulation boxes from each $\xion$ bin. If the bin has $< 25 $ simulations then we take all the boxes in that bin. This leaves us with the more balanced distribution shown in the bottom panel of \autoref{fig:unbal_dist}. This is the same methodology as used in \citet{Hassan:2019}.
     \item Once we have a balanced training set, we sample different 2D slices from our simulation boxes. \cite{Hassan:2019} stated that, based on their experience, a separation of $\sim 4$ Mpc produces sufficiently independent 21~cm maps. Given the dimensions of our simulation boxes this results in 40 unique 2D slices along each axis. We sample 40 slices from each of the $x$-, $y$- and $z$-axes of our boxes, randomly choosing the position of the first slice. We set peculiar velocities to be zero, allowing us use of all three dimensions, maximising our dataset. Ideally, one would simulate enough boxes to retain the peculiar velocity element or even better apply the architecture to lightcones, however, this is reserved for future work and this simplification follows the literature we are surveying.
     
    \item
    To train the network, each slice needs to have an associated label which indicates the target parameters. This is different for each of our case studies. For this case study, we calculate the $\xion$ fraction for each of the slices in our dataset and use that as the training label. To train a network efficiently, the labels have to be standardised. The weights in the network are adjusted during training, and their values are affected by the depth of the network (i.e., the number of layers). Since the weights can change exponentially with depth, unstandardised labels could lead to the multiplication of large weights and numerical issues such as overflow (values becoming too large) or underflow (values becoming too small). In order to standardise each parameter, we subtract the mean for that parameter calculated over the training set and divide by the standard deviation.
    
    \item 
    The final step is to separate the 2D slices into training, validation and test sets. Once the 40 2D slices were sampled along the $x$-, $y$ and $z$-axes, we randomly allocated 32 slices to the training set, four slices to the validation set and the remaining four slices to the test set for a training/validation/test split of 80/10/10. Note we are including data from the same box in the different data sets not as a design choice, but in order to follow the literature and test the limitations of this approach.
    
\end{enumerate}

\subsection{Training and results}

We train our network in the usual way by minimising the average loss function, where the loss function measures how well the neural network is performing on a single training example. The optimised weights after each epoch are used as the starting point for the next epoch. In this way, the average loss function is gradually reduced as the network becomes better at predicting the target parameters. At the end, the weights that produce the smallest loss are taken for the final network.

\begin{figure}
\centering
\includegraphics[width = \columnwidth]{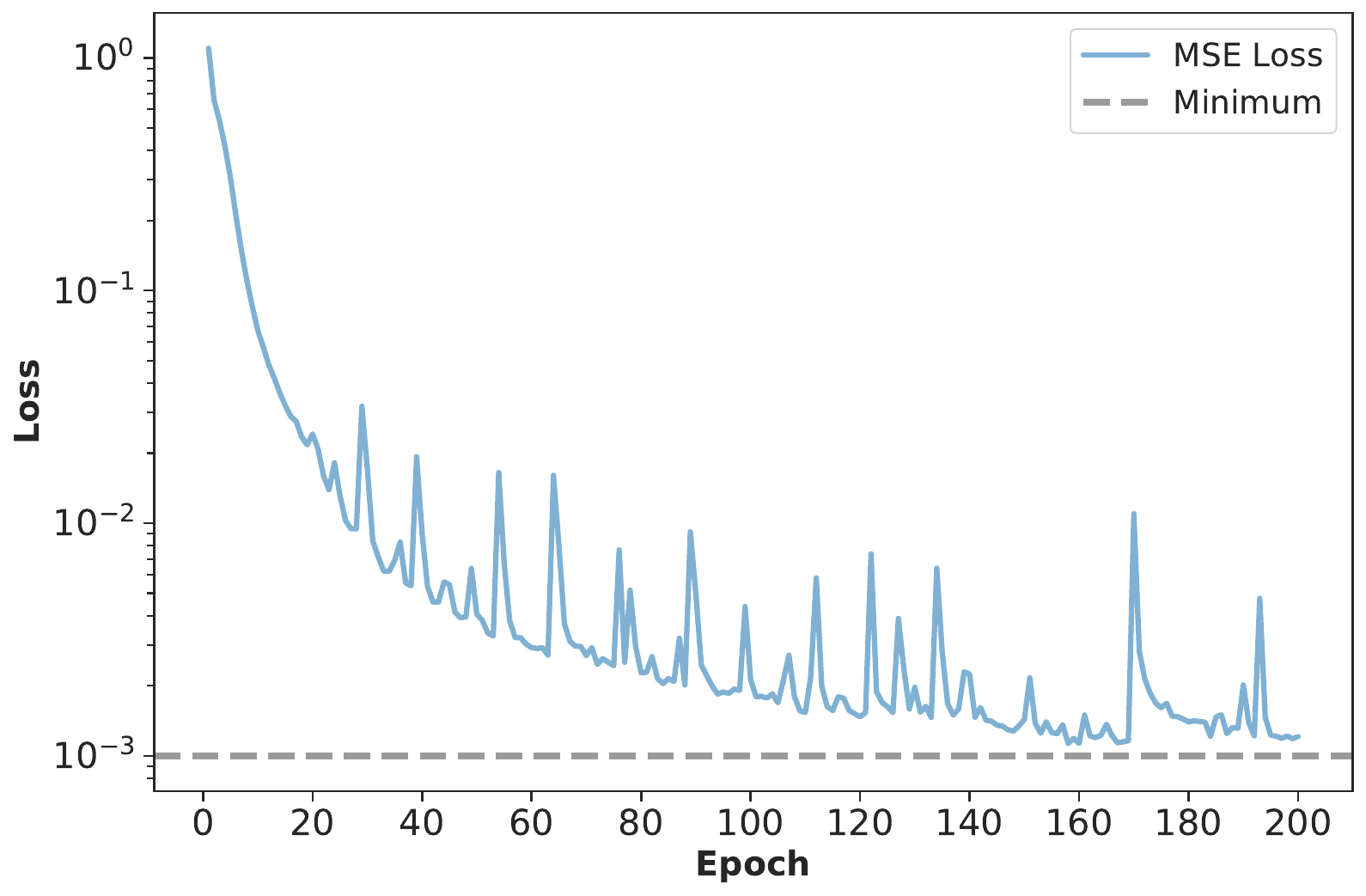}
\caption{The network loss for Case 2 as a function of training epoch. This is representative of similar plots for the other case studies. We can see that the loss decreases sharply initially, then tapers out as the network learns. The loss function appears to be approaching an asymptotic minimum of $\approx 0.001$. In each case study, the network with the minimum loss was used for the remainder of the analysis.}
\label{fig:lossVepoch}
\end{figure}

An example plot of the loss over the training epochs is shown in \autoref{fig:lossVepoch}, where the network was trained for 200 epochs using the parameters shown in Table 2. We can see from this figure that the loss decreases sharply initially and then flattens, reaching a loss of $\sim \!10^{-3}$ at epoch $ = 200$. Although further training might improve the network performance, beyond this point there is a diminishing return. For all the case studies in this paper, we see similar behaviour and so train for 200 epochs and make use of the network with lowest loss.

\begin{figure}
\centering
\includegraphics[width = \columnwidth]{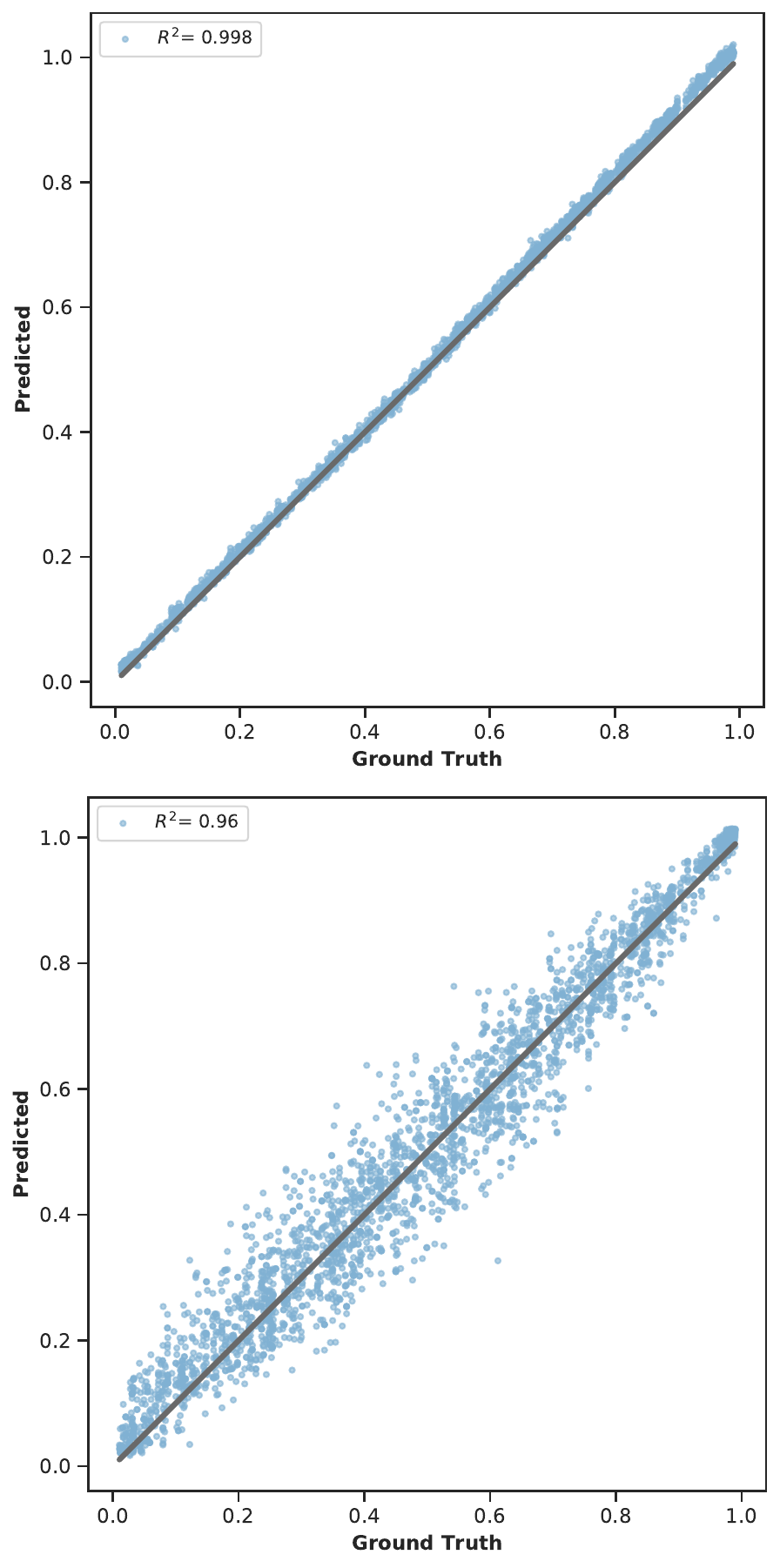}
\caption{Predicted ionized fraction against true ionization fraction for the case where testing is on slices from boxes represented in the training set (top panel) and on slices from newly simulated boxes (bottom panel). While $\xion$ is still recovered to the degree that $R^2$ is only slightly reduced, the increased scatter in the bottom panel indicates that the network may not be fully generalised, and is instead learning the specific boxes shared in training and testing.}
\label{fig:mangena_result_1}
\end{figure}

The results from this network are shown in \autoref{fig:mangena_result_1} (top panel).  For the majority of the points the recovered values are very close to the true values and the correlation coefficient is $R^2=0.988$, indicating very close agreement. This indicates that the network has successfully learnt how to predict $\xion$ from a $21$ cm image.

\subsection{Testing the generalisation of the network}
\label{sec:gen_test}

One goal of this work is to test how the network will perform when trained on one set of simulations and tested on another. To that end, we test how the network performs on a simulation box that it has not seen. In the previous section we trained and tested the network on different slices from many simulation boxes, each with its own random initial conditions. While the same slices are not repeated in the training and test sets, the same boxes are represented in both. Recall from \autoref{sec:preproc_1} that random slices were taken from each box, and these slices were divided into the training, validation and test sets. Therefore slices from each box are present in all three sets. 

To test how well the network is generalised to unknown boxes, we simulate new boxes and test the network on those. While the parameter we are predicting will be in the same range as the training set, the initial density field used to generate these new boxes is completely different from any of those used in the training dataset.

The results of testing the network on slices from newly simulated boxes is shown in \autoref{fig:mangena_result_1} (bottom panel). The $R^2$ value has only slightly reduced, indicating an excellent recovery of the $\xion$ parameter. However, there is clearly an increased scatter, indicating the network has previously been using information from slices sharing the same density field to make the predictions in the top panel, and not purely learning the underlying physics. For application to real $21$ cm observations there is no way to guarantee that the true physical density field is perfectly replicated in the training set, and so generalising the network as in the bottom panel is a more robust approach. For this parameter, the effect is not enough to seriously endanger the recovery, however for other parameters this may not be the case, motivating the next sections.



\section{Adding Complexity with more Cosmological and Astrophysical Parameters}
\label{sec:hassan_et_al}

In the previous case study, we looked at inferring only the ionization fraction. Future 21~cm experiments will, of course, be interested in inferring more information about the underlying astrophysics. \cite{Hassan:2019} extended their work in \cite{Mangena:2020} to look at using CNNs to infer six parameters describing a mix of cosmology and astrophysics from each individual 21~cm image. The underlying framework of both \cite{Hassan:2019} and \cite{Mangena:2020} are fundamentally the same, with the two papers primarily differing in the target number of parameters. By increasing the number of target parameters, we expose the network to more intricate relationships within the data. This forces the network to focus on patterns that are genuinely representative of the underlying processes, rather than overfitting to superficial patterns in the specific simulation slices making up the training set.

If the network performs well on unseen data with a larger number of target parameters, it is more likely to have learned the true underlying features of the data and be truly generalisable.

\subsection{Data pre-processing}
\label{sec:preproc}

To produce our dataset for this case study, the same pre-processing steps as in \autoref{sec:mangena_et_al} are followed, except that in Step~(iv) we now label each slice with six parameters rather than simply the ionization fraction. This allows the use of CNNs to produce a mapping from the single input slice to each of the six inferred parameters.

\subsection{Training and results}

\begin{figure*}
\centering
\includegraphics[width = \linewidth]{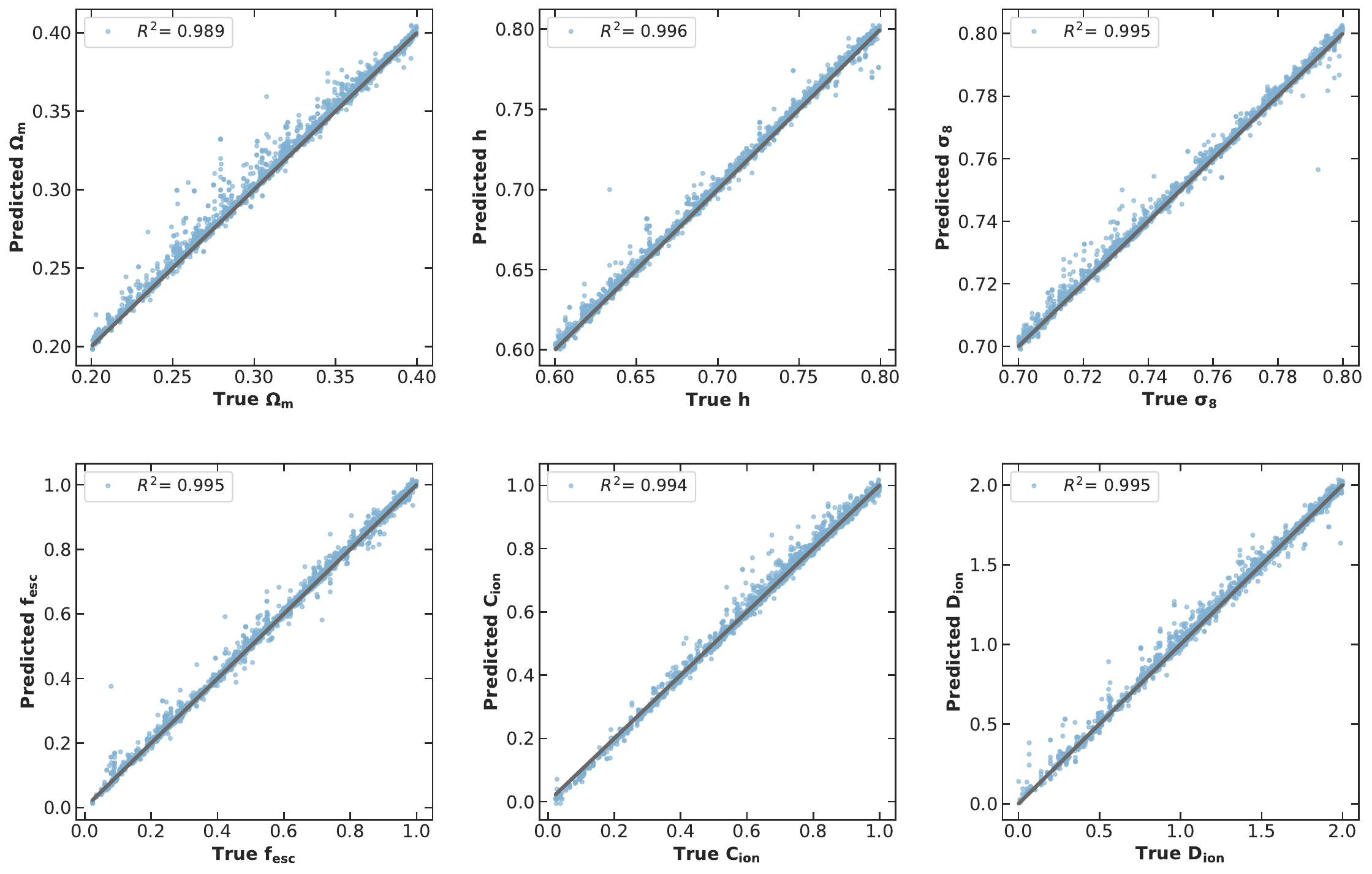}
\caption{Plot showing the results of training the network on the dataset described in section \autoref{sec:preproc}. Slices from each simulation box were sampled and put into the training set, then different, uncorrelated slices, were sampled and put into the testing set. These plots show how the network performs on the testing set over the six parameters. It can be seen that the network performs very well across all parameters with an average $R^2 \simeq 0.994$.}
\label{fig:hassan_result}
\end{figure*}

We first reproduce the results presented in \cite{Hassan:2019} using our network. Again we train for 200 epochs, which achieves a similar cost function value to \autoref{sec:preproc_1}. We use the same training, validation and test sets as in \autoref{sec:preproc_1}, although the labelling is changed to reflect the new target parameters. The results of applying the network to the test set are shown in \autoref{fig:hassan_result}. As expected, with an average $R^2 \simeq 0.994$, the network can be seen to match the performance of the same network in \cite{Hassan:2019}, which found an average $R^2 \simeq 0.992$. For each parameter, there is a clear correlation  between input and output parameter with only minimal scatter.

\begin{figure*}
\centering
\includegraphics[width = \linewidth]{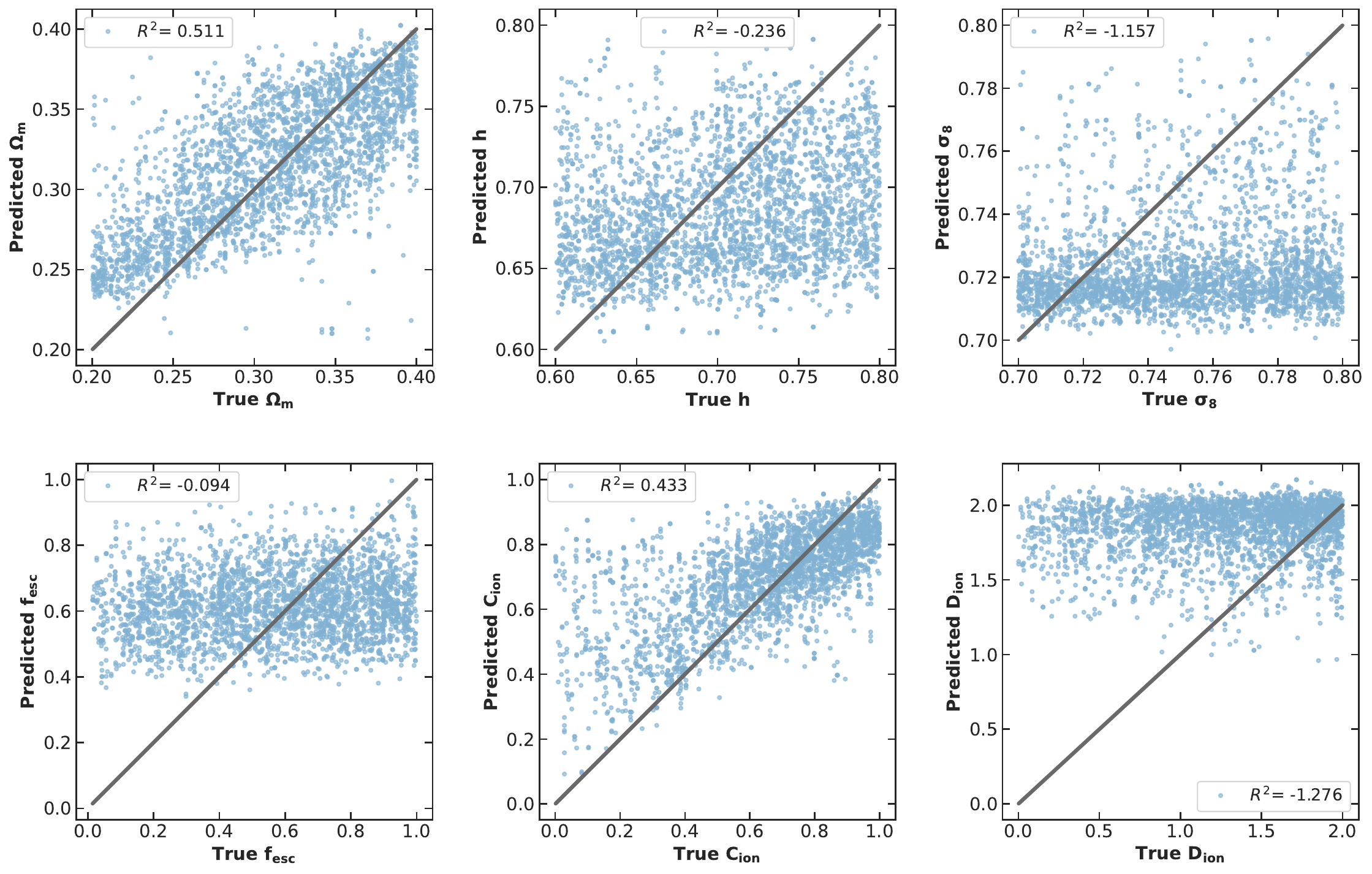}
\caption{Plot showing results of training the network on the dataset described in section \autoref{sec:preproc}. Unlike in \autoref{fig:hassan_result}, slices for the testing set were sampled from completely different simulation boxes to the training set. These plots show how the network performs on the testing set over the six parameters. It can now be seen that the network performs poorly across the six parameters. This indicates that in the previous case, the network was somehow exploiting an association of slices to a specific box rather than learning the underlying physical relationships producing features of the brightness temperature.}
\label{fig:gen_test}
\end{figure*}

As in the previous case study, we now examine the ability of the network to generalise by testing on a new set of simulation boxes, unseen by the network. The results are shown in \autoref{fig:gen_test}. We can see that there is little correlation between the ground truth and predicted values for four out of six of the parameters, namely ($h,\sigma_8,f_{esc},D_{ion}$). Even for the two parameters with signs of correlation ($\Omega_m,C_{ion}$), there is a large amount of scatter. Therefore, we can conclude that the network performs very poorly when tested on boxes that were not present in the training set.

In \autoref{sec:mangena_et_al} it appeared that the network successfully learnt how to predict the ionization fraction from an input brightness temperature map and that this generalised to previously unseen maps. In this section, we have seen that the network seemed to successfully predict six parameters testing more complex relationships, though using slices taken from boxes also used in the training set. However, the network failed to predict the six parameters when given a slice from newly simulated boxes. This suggests that the network has not robustly identified features correlated with the parameters, but has instead learnt to associate slices with individual boxes from the training set. Rather than inferring the parameters directly, these results hint that the network is identifying a specific box and then returning the parameters of that box. Such a learnt approach is highly unlikely to generalise to being able to infer parameters from real data.

\section{Quantifying the effect of shared training-testing boxes}
\label{sec:5.3}
In the previous sections, we showed that the network appears to make excellent predictions when it has a large number of slices from simulation boxes shared between training and testing. These predictions falter when it has a very low number of slices from shared simulation boxes. This suggests that the network is not generalising, but rather learning specific features of each box. For example, when provided with multiple slices from a single box, the network may be learning to identify the shared origin of the slices via the shared underlying density field. 

We now want to test the correlation between the number of slices used from each box in the training set and the performance of the network on other slices from those boxes. To this end, we generate a series of networks each with a different number of slices per box used for training:
\begin{itemize}
    \item 
    The data is simulated as outlined in Steps 1 and 2 from \autoref{sec:preproc_1}.
    \item 
    For each ionization bin, we sample 120 slices, 4 Mpc apart, from each box. This spacing is chosen so that different slices are spatially uncorrelated.
    \item 
    $P$ slices per box are randomly selected to include in the training set. The remaining $120-P$ slices from each box are used for the testing set. This is repeated for different values of $P$ ranging from $0$ to $105$ to construct different training and testing sets to train and test different copies of the network.
    \item 
    For a fixed number of boxes varying $P$ varies the size of the training set. But we would like to ensure that each time we train a network the training sets used are comparable in size. To achieve this, we randomly select a subset of the boxes to include slices from in the training set. The number included is chosen to ensure that the product of $P$ and the number of boxes sampled from is roughly constant. This allows us to take exactly $P$ slices per box, which we could not do, for example, by selecting a fixed number of slices from the full candidate pool. 
    \item 
    The testing set likewise differs in size for different values of $P$. This is less important to normalise over since we expect even the smallest testing set to be representative.
    \item 
    This process is repeated for each of the 10 ionization fraction bins.
    \item 
    For each value of $P$ we retrain the network using the appropriate training/testing sets.
\end{itemize}

\begin{figure}
\centering
\includegraphics[width = \linewidth]{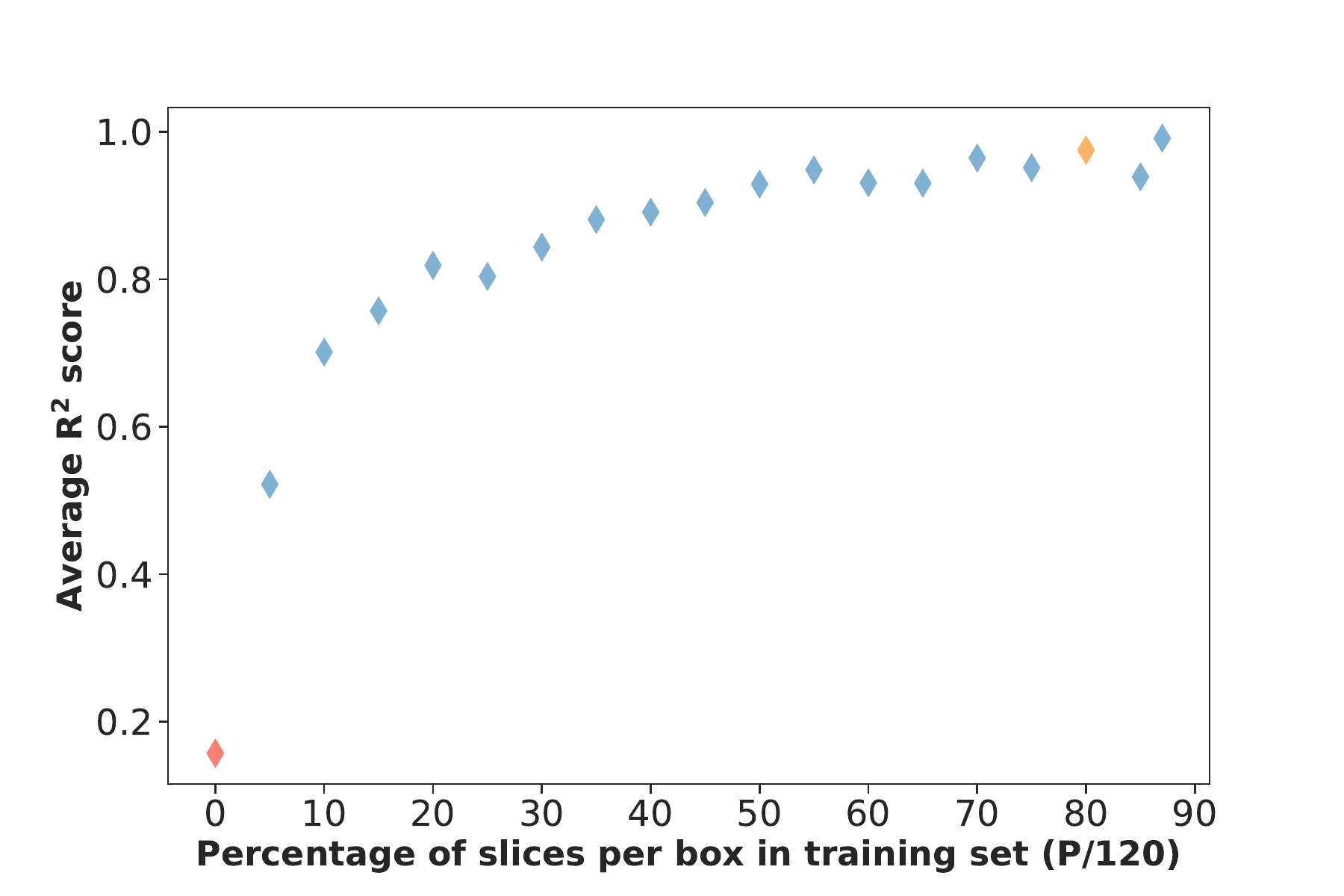}
\caption{The average $R^2$ score as a function of the fraction of slices in the training set. The orange point corresponds to the case of \autoref{fig:hassan_result} and the red point corresponds to the case of \autoref{fig:gen_test}. 
}
\label{fig:avg_r2}
\end{figure}

We create training sets for 18 different values of $P$. We train and then test a network on each of these training sets. For each, the $R^2$ score was calculated for each of the six parameters and then an average was taken. The average $R^2$ score for each group is shown in \autoref{fig:avg_r2}, where the $x$-axis here is $P/120$,  the fraction of repeated sampling from each individual box. For large $P$, the network `sees' many slices from each box and so has the potential to learn how to identify that box specifically. For small $P$, the network sees only a handful of slices from each box and struggles to correctly infer the underlying parameters.   

We can see from \autoref{fig:avg_r2} that the results agree with the conclusion of the previous subsection. As the number of training slices from a box shared in testing is increased, the performance of the network increases. This shows that the network is not learning the underlying physical relationships and not generalisable to unknown boxes. As it stands now, the network is not general enough to perform well on real data.

\section{Including Redshift Structure into the Network: Recovering mid-point and duration of reionization}
\label{sec:laplante_et_al}

In this section, we will use CNNs to infer the duration and mid-point of reionization from brightness temperature slices. Previous research \citep{LaPlante:2018} found that the CNNs could infer the mid-point of reionization to a precision of 1\% and the duration of reionization to a precision of 5\%.



\subsection{Data pre-processing}
\label{sec:preproc_2}

We must first post-process our simulations to calculate the ionization history parameters, such that we can use them as labels. We use the same labels as in \citet{LaPlante:2018}. These are the mid-point of reionization $\zmid$, defined as the redshift at which the simulation box is 50\% ionised, and the duration of reionization $\Delta z$, taken to be the difference between the redshifts at which the simulations box is 75\% and 25\% ionised.

To get well-defined ionization history parameters, we need to ensure that these three ionization fractions are present in our simulation outputs, which was achieved with these pre-processing steps:

\begin{enumerate}
    \item 
    We check that the difference in ionization fraction between the lowest and highest redshift is greater than of equal to 0.75. If the simulation does not contain this range of ionization fractions we discard it. This ensures we capture the range $x_{\rm HII}=0.25$ to $x_{\rm HII}=0.75$ in our simulation outputs. This criterea captures more of the ionization history than is strictly necessary here, but the extra range captured will be useful for the fourth case study in \autoref{sec:billings_et_al}. 
    \item Once we have discarded all simulations that fail the previous step, for each simulation we draw at random one 2D slice from each of the 20 redshift boxes and combine them to provide a single (200, 200, 20) datacube. 
    \item $\Delta z$ and $\zmid$ are calculated for each datacube and used as labels when training the network. Finally, the dataset is split into a training/testing split of 80/20. 
\end{enumerate}

At the end of this, we have reduced an initial set of $30,000$ simulations to the $\sim\!20,000$ datacubes that form our training/testing set. Each datacube has known labels in terms of $\zmid$ and $\Delta z$, and represents a series of twenty $21$ cm maps with known redshifts. This additional redshift structure represents the key difference between the network approach here and that of \autoref{sec:mangena_et_al} and \autoref{sec:hassan_et_al}.


\subsection{Training and results}

\begin{figure}
\centering
\includegraphics[width = \linewidth]{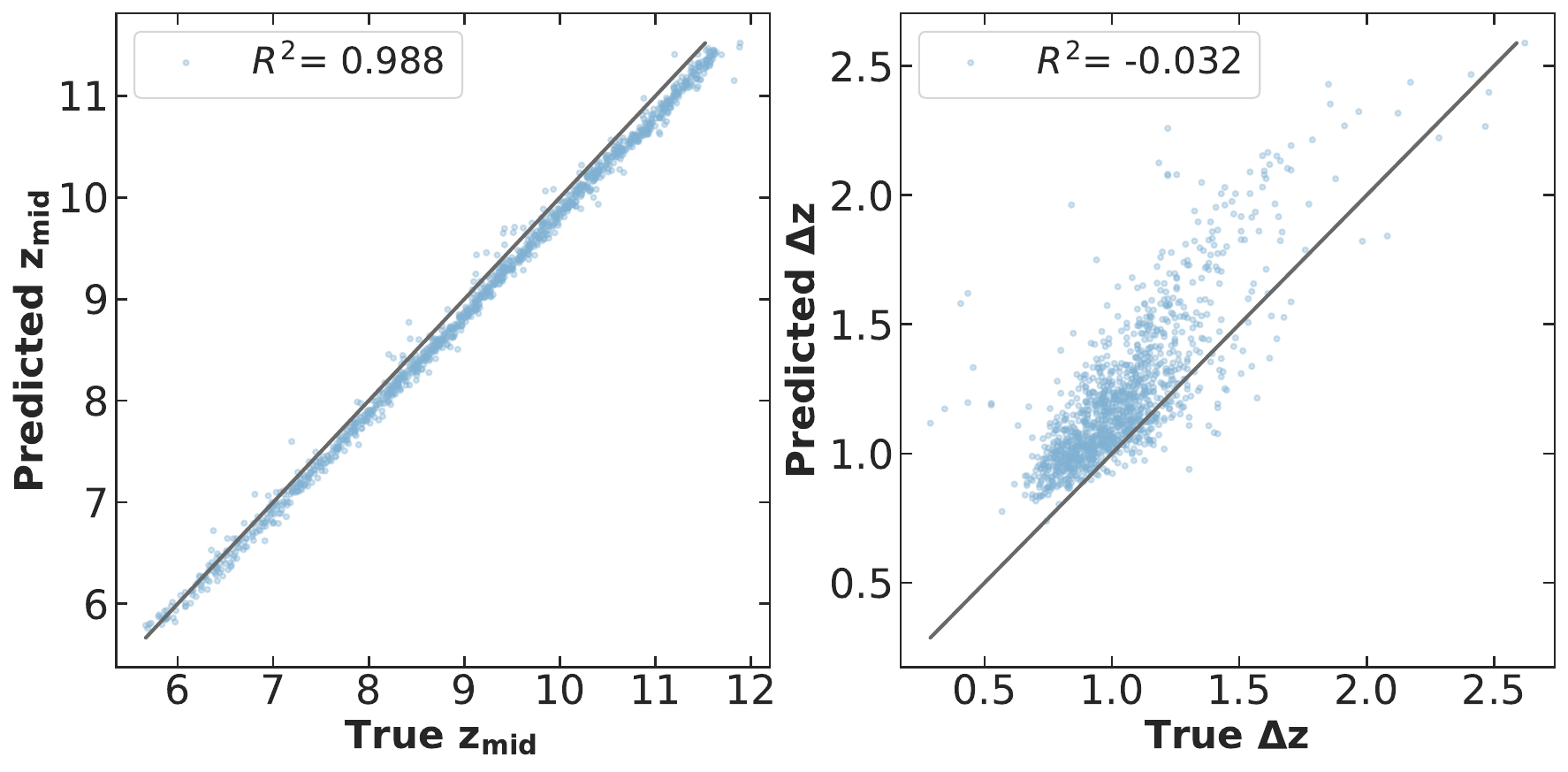}
\caption{Plot showing results of training network on 3D SimFast21 data with $\zmid$ (left) or $\Delta z$ (right) labels. It can be seen that the network performs well for $\zmid$ with a $R^2 \simeq 0.988$ and significantly less well for $\Delta z$. 
}
\label{fig:zmid_deltaz}
\end{figure}

The network was trained for 200 epochs using the parameters shown in \autoref{tab:train_param} as before. We trained separate networks for $\zmid$ and $\Delta z$. The results for each of these are shown in \autoref{fig:zmid_deltaz}. In the left panel of \autoref{fig:zmid_deltaz}, we see that the network does a fairly good job of predicting the true value of $\zmid$, although we see that it has a tendency to under-predict the true value. In contrast, the second network struggles to predict $\Delta z$.

The failure of the network to predict $\Delta z$ is a consequence of the limited range of sampling of that parameter in the training set. Looking at the lower panel of \autoref{fig:zmid_deltaz}, we can see that most of the points are clustered around the middle. Recall that the training set has been created by uniform sampling of the astrophysical parameters followed by post-processing to ensure the labels $\zmid$ and $\Delta z$ were well-defined. This procedure is typical of a simulation first approach for generating a dataset, but is in no way guaranteed to produce good sampling of derived parameters.

\begin{figure}
\centering
\includegraphics[width = \linewidth]{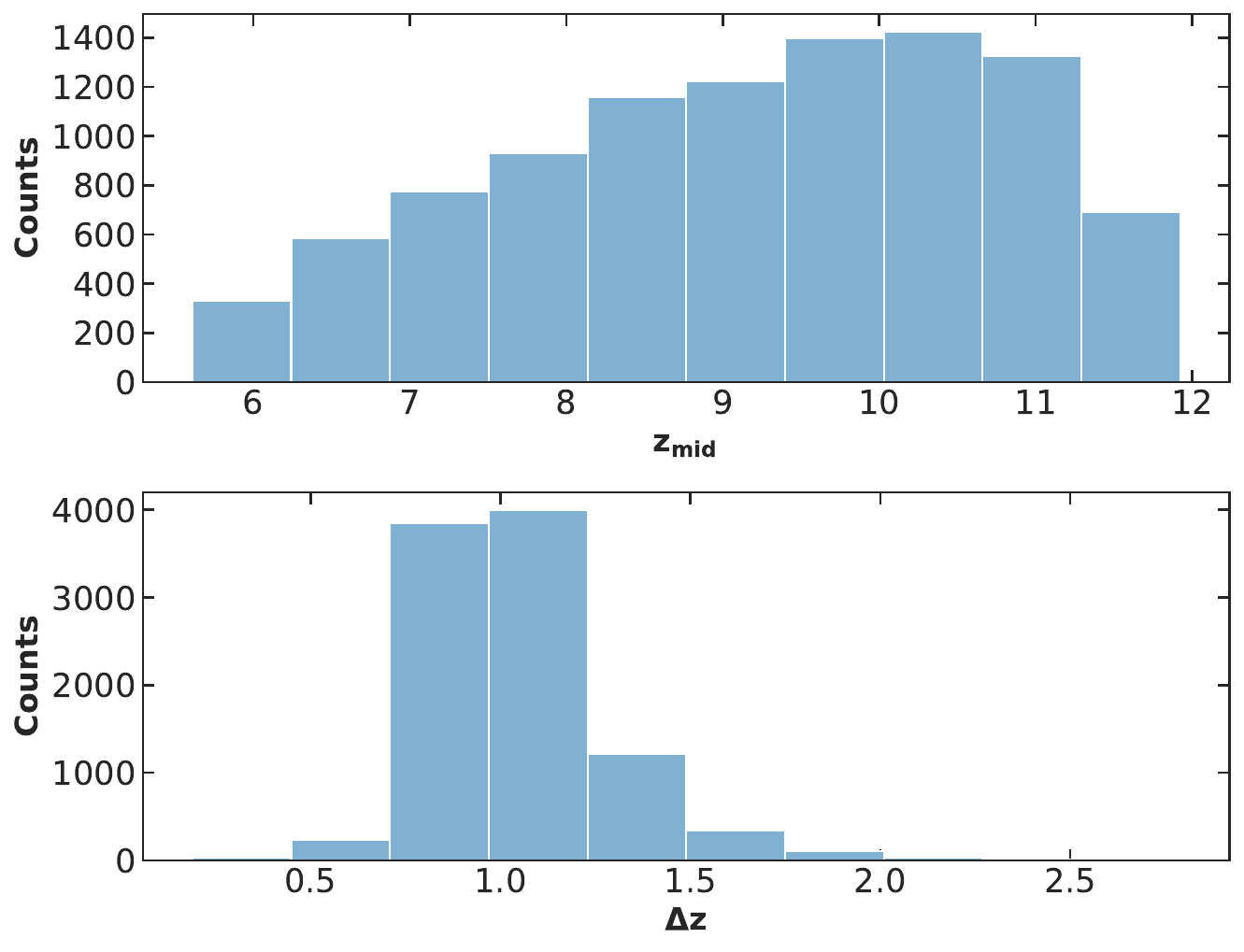}
\caption{Distribution of the $\zmid$ (top panel) and $\Delta z$ (bottom panel) parameter values in the training set. 
}
\label{fig:zmid_deltaz_hist}
\end{figure}

\autoref{fig:zmid_deltaz_hist} shows the measured distributions of the two derived parameters. It indicates that $\zmid$ is well sampled, although higher values are somewhat over-represented, but only a narrow range of $\Delta z$ values are captured in the simulations. The post-processing described in \autoref{sec:preproc_2} limits us to the samples of $\Delta z$ that fall within the middle of our simulations, leading to the distribution we see in \autoref{fig:zmid_deltaz_hist}. 

This effect is due to "out-of-distribution" (OOD) samples, data points that fall outside the range of the data distribution on which a machine learning model was trained, and can heavily influence the results of the network, leading to poor predictions \citep[e.g.,][]{hassan2024}. It is therefore vital to carefully consider the statistics of the training set. In \autoref{sec:hassan_et_al}, this meant looking at the correlation between slices and the distribution of ionisation fractions. In this case, this means ensuring that the distribution of parameters in the training set gives a good covering of the representation values that the model needs. Failure to do this can bias the network to be very good at identifying over-present training values and effectively expect those values more often than reasonable in unseen data. 



\section{Including Redshift Structure into the Network: Recovering CMB Optical Depth}
\label{sec:billings_et_al}

We now turn to recovering the CMB optical depth, $\tau$. Previous research has suggested that CNNs can infer the CMB optical depth to within $3\%$ \citep{Billings:2021}. However, \citet{Zhou_2022} found that $\tau$ is sensitive to the semi-numeric reionization scheme used, and a CNN trained on one semi-numeric code does not generalise well when tested using simulations from another semi-numeric code. In this section, we will pre-process our data so that we end up with a dataset that is similar to that of  \cite{Billings:2021}.


\subsection{Data pre-processing}
\label{sec:preproc_billings}

Our pre-processing here broadly follows the procedure described in \autoref{sec:preproc_2}, except that we use the CMB optical depth as the label for the datacubes. 



As in \autoref{sec:preproc_2} we remove those boxes that do not contain a change in ionization fraction, $\xion$ of at least 0.75 across the full redshift range. This criterion ensures that enough of the reionization history is present to give a sensible numerical estimate of $\tau$. Even so, for some cases the estimate of the optical depth from the limited range in the simulation will differ from the true optical depth calculated from the full simulation history; this is not ideal but is not necessarily a problem, as the network will learn on the basis of whatever label it is given. It does, however, mean that the $\tau$ label is not identical to the true CMB optical depth and so caution should be taken in its interpretation. Empirically, this ambiguity seems to have little impact on how well the network is able to predict the value of $\tau$.

\subsection{Training and results}

\begin{figure*}
\centering
\includegraphics[width = \linewidth]{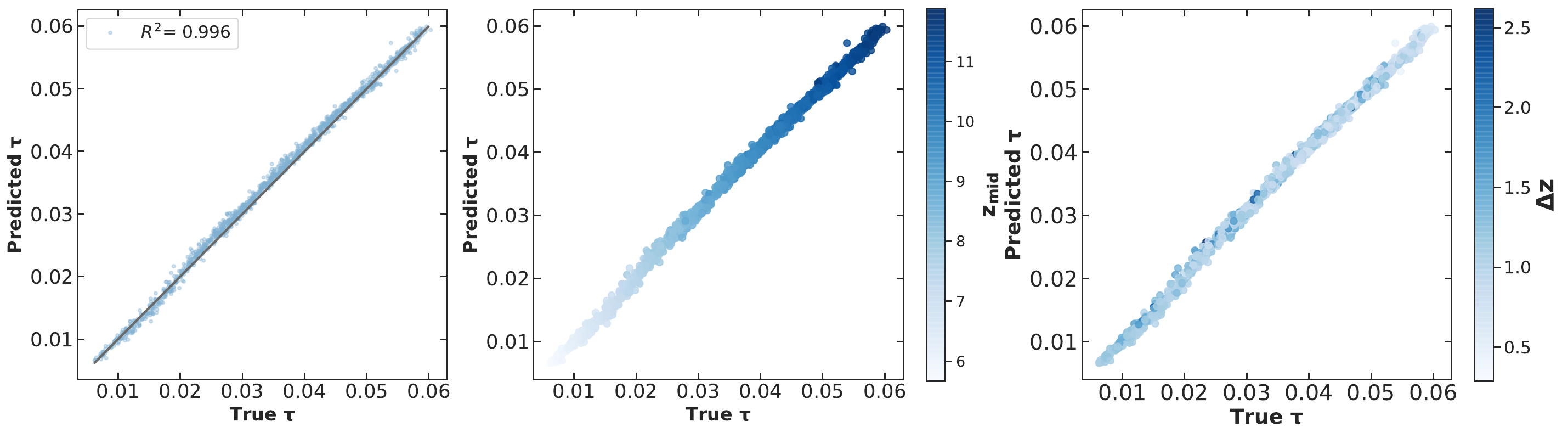}
\caption{Left panel: The results of training network on 3D \texttt{SimFast21} data with $\tau$ labels. Middle and right panels: Results of training network on 3D SimFast21 data with $\tau$ label. Each point is coloured according to the value of its $\zmid$ (middle) or $\Delta z$ (right) label.
}
\label{fig:tau}
\end{figure*}

The network was trained for 200 epochs using the parameters shown in Table 2 as before. The resulting inferences of $\tau$ on the test set are shown in the left panel of \autoref{fig:tau}. The network does well at inferring the optical depth. This is consistent with the results from \autoref{sec:laplante_et_al}, where the network did a good job of predicting $\zmid$, which is strongly correlated with $\tau$; the fact that it could not predict $\Delta z$ is unimportant as $\tau$ is (almost) independent of the duration of reionization. 


As we have the ground truth values for all parameters we can combine the labelling of the previous section with the results of this section to further test these relationships.  As shown by the symbol colours in the middle and right panels of \autoref{fig:tau}, there is indeed a strong correlation between $\tau$ and $\zmid$ (shown by the obvious colour gradient) and no significant correlation between $\tau$ and $\Delta z$. 


In summary, the inference of $\tau$ generalises well, but for the $(\zmid, \Delta z)$ parameterization the inference is poor. This compares to the results of \autoref{sec:hassan_et_al}, where seemingly accurate reconstructions hid a subtle correlation between simulations via a shared density field. These two very different situations both indicate the sort of subtleties involved in creating a suitable training sets for 21 cm cosmology.  It appears that parameters like $\tau$ or the ionization history $\xion$ are easier for networks to predict, compared to astrophysical parameters like $f_{\rm esc}$. This is potentially an indicator that details of the morphology are harder to capture in the CNN response. 


\section{Synthesis of approaches}
\label{sec:sixparams}

The techniques described above can be combined by using our training datacubes described in \autoref{sec:laplante_et_al}, which incorporate 2D slices with known redshift information, but now labelling each datacube with all 6 astrophysical and cosmological parameters, as in \autoref{sec:hassan_et_al}. This explores the question of whether including the extra redshift information allows the network to generalise in the way it failed to previously when attempting to recover the astrophysical and cosmological parameters.

\begin{figure*}
\centering
\includegraphics[width = \linewidth]{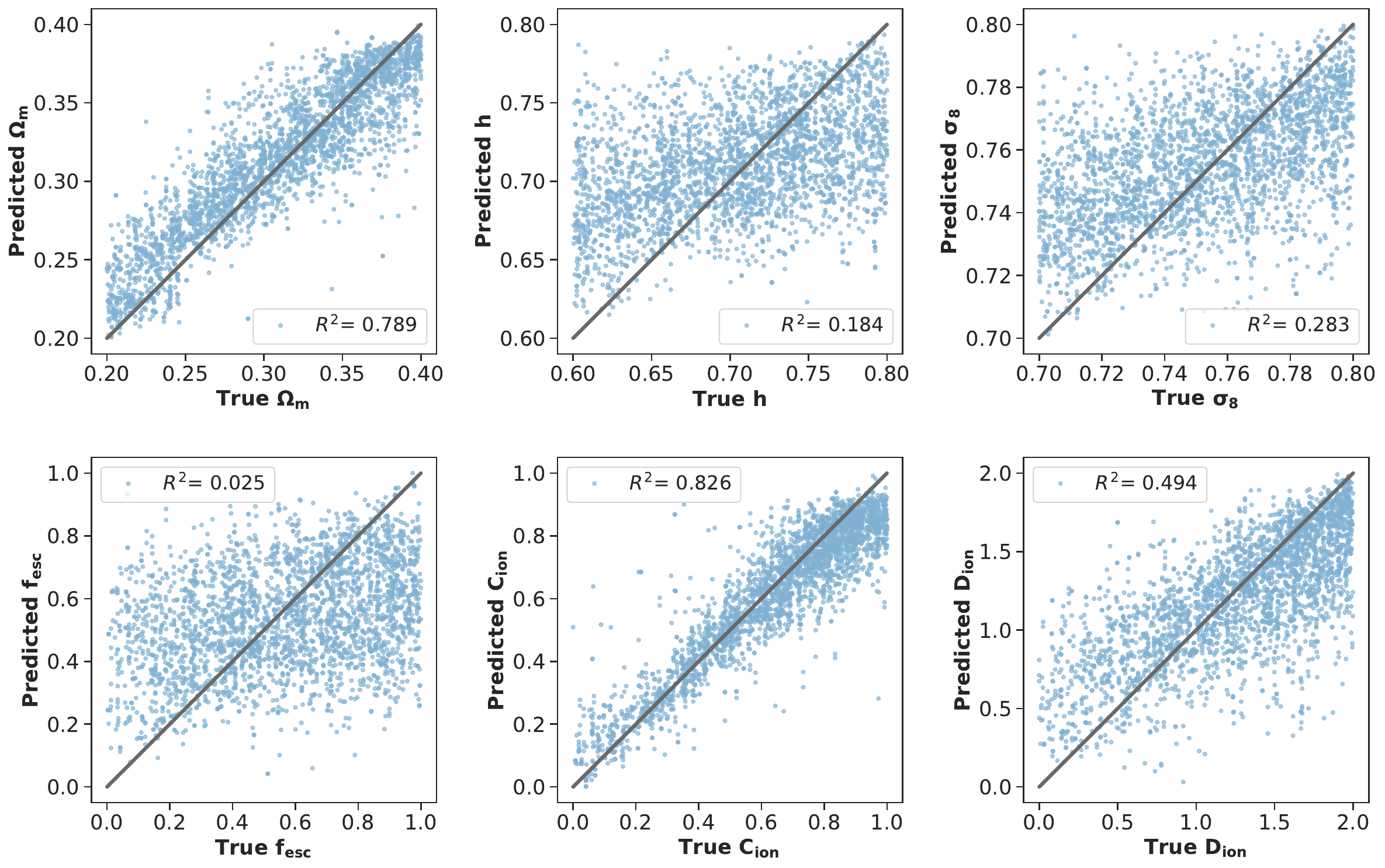}
\caption{The results of training the network on the new 3D dataset.}
\label{fig:hparam_20b}
\end{figure*}

These results of this exercise are shown in \autoref{fig:hparam_20b}. This parallels our previous results in \autoref{fig:gen_test}, but the training data now includes the additional redshift structure. The network is now able to infer four out of the six parameters: $\Omega_m$; $\sigma_8$; $\Cion$; and $\Dion$. This is a significant improvement relative to \autoref{sec:hassan_et_al}, in which one or two parameters were able to be inferred and even then with broader scatter in the reconstructions. 

This improvement is promising for methods of this type, particularly because the different redshift slices for a single datacube are selected randomly from different \texttt{SimFast21} realisations, so the network sees the average time evolution of the field, but not the time evolution of a specific density region. So the network must be learning something about the nature of the evolution of the ionization maps with redshift.  Further work will be needed to fully understand the way in which the network is learning, but this is a promising indication that given the right inputs that ML methods can be effective in 21~cm cosmology.

\section{Conclusions}
\label{sec:conc}

We have explored how robustly CNNs can infer astrophysical and cosmological parameters from 21~cm maps by reproducing and testing the generalisability of networks in use in the field \citep{Mangena:2020,Hassan:2019,LaPlante:2018,Billings:2021}. This highlights some of the challenges in building a representative training set and the failure of trained networks to generalise beyond the datasets used to train them. We then combine some of the lessons learnt from these case studies to build a (somewhat) more generalisable network in \autoref{sec:sixparams}.

In our first two case studies, a CNN is applied to individual 21cm maps to predict the neutral fraction (\autoref{sec:mangena_et_al}) and astrophysical parameters (\autoref{sec:hassan_et_al}). We found that the training sets could give the network multiple slices from a single box in a way that may allow the network to train by identifying the features of a single box rather than the features of the reionization process. Removing this ability by limiting the training set to a single slice per box significantly degrades the ability of the network to infer astrophysical parameters. More fundamentally, the network struggles to generalise what it has learnt to infer astrophysical parameters from images from newly simulated boxes that are not represented in the training set. We investigated how this generalisability depended on the fraction of slices from a shared box and found, as expected, that as the number of training slices from a
box shared in testing is increased, the performance of the network increases. This indicates that the network is not useful as it stands for application on real data, where we will have trained on a realisation of an unknown density field (i.e. $P=0$).

The second pair of case studies looked at cases where the CNN is given a set of images at known and increasing redshifts and is asked to predict the ionization history (\autoref{sec:laplante_et_al}) or the CMB optical depth (\autoref{sec:billings_et_al}) following the approaches of \cite{LaPlante:2018} and \cite{Billings:2021}, respectively. Here our training set is derived from simulations that do not allow the a priori labelling of data cubes with reionization mid-point, duration, and optical depth. Instead, these labels have to be derived from the simulation outputs, which constrains our ability to uniformly sample these parameters for the training set. The poor sampling of the reionization duration translates into poor performance of the network for inferring the reionization duration, despite its otherwise good performance for the optical depth and the midpoint of reionization.

This work highlights some of the limitations of CNNs as an inference tool. We have seen that even when testing seems to indicate very good performance from the CNN that this can be misleading. The networks can learn the wrong lesson from the input training set leaving their ability to infer parameters fragile and dependent on artificial properties of the inputs, for example as in \autoref{sec:hassan_et_al} when testing and training sets were composed of independent slices drawn from the same simulation boxes. Successful implementation on one simulation code may not generalise to data from a second simulation code \citep[see also][]{2022PASP..134d4001Z}. More generally, sufficiently sampling parameters that must be derived from simulations rather than specified directly can lead to very limited performance from the networks. CNNs and deep learning may have the power to transform analysis of future 21cm data and cosmological images, but their black box functionality places a high level of responsibility on the user to ensure that they are being applied correctly. 

This generic problem of handling out-of-distribution samples, since the Universe will always be imperfectly modelled by our simulations, is increasingly being recognised and we hope to explore this further in future work \citep{2023arXiv231118007G}.


\section*{Acknowledgements}

KS acknowledges funding by an Imperial College President's Scholarship. EC acknowledges the support of a Royal Society Dorothy Hodgkin Fellowship and a Royal Society Enhancement Award. LC acknowledges the support of STFC consolidated grant ST/X000982/1.

KS would like to thank Sultan Hassan for many useful conversations and for providing the training sets from previous work. The authors would also like to thank Paul La Plante for his useful comments on the paper.

\section*{Data Availability}

The data underlying this article will be shared on reasonable request to the corresponding author. The code underlying this article is available at \url{https://github.com/kimeels/21cm}.




\bibliographystyle{mnras}
\bibliography{main} 

%

\label{lastpage}
\end{document}